\documentstyle[aps,preprint]{revtex}
\def\be{\begin{equation}}
\def\ee{\end{equation}}
\def\bea{\begin{eqnarray}}
\def\eea{\end{eqnarray}}

\begin{document}
\draft

%%%%%%%%%%%%%%%%%%%%%%%%%%%The Manuscript%%%%%%%%%%%%%%%%%%%%%%%%%%%%%

\preprint{CERN-TH/98-30 }

\title {\bf HIDDEN SYMMETRIES OF M THEORY }

\bigskip

\author{ Jnanadeva Maharana \footnote{e-mail: maharana@nxth04.cern.ch \\
Permanent address: Institute of Physics, Bhubaneswar 751005, India} }
\address{CERN\\
CH-1211 Geneva 23, Switzerland 
}
\maketitle

\begin{abstract}
A worldvolume action for membrane is considered to study the target space local
symmetries. We introduce a set of generators of canonical transformations to
exhibit the target space symmetries such as the 
general coordinate transformation
and the gauge transformation of antisymmetric tensor field. Similar results are
derived for type IIB string with manifestly S-duality-invariant worldsheet
action.

\end{abstract}
\vspace{2 cm}

\noindent CERN-TH/98-30 \\
January 1998
\narrowtext

\newpage

%\section {Introduction}
%\label{int}

There has been  considerable progress  in our understanding of
nonperturbative aspects of  string theory \cite{asen,w}.  
It is recognized that 
 extended objects, such as the p-branes and D-branes,
have played a key  role in these developments \cite{jp}. They    
  appear as non-perturbative
solutions of the low energy string effective action  and they  
have been instrumental for our understanding of duality symmetry 
conjectures in
string theory and in providing insights into string dynamics in various 
dimensions.  
 It is now accepted that there are intimate connections amongst  the five 
string theories 
and that there is an underlying fundamental theory, M-theory or F-theory,  
and that the five different 
string theories are
manifestations of various
phases of that  theory \cite{jhs,pk1}. 
It is also believed that the low energy effective action
of the M-theory can be identified with that of  $D=11$ supergravity.  There is 
mounting evidence that M-theory encompasses and unifies string theories and  
string dynamics in diverse dimensions. 
 The low energy effective action
of this
theory contains the antisymmetric tensor, besides the graviton, in its bosonic
sector. Therefore, the membrane that  couples to the three index 
antisymmetric tensor field is expected to be the natural extended fundamental 
object \cite{bts} in eleven dimensions,  
 with five-brane as 
its solitonic counterpart; consequently,  a 
lot of  attention has been focused on the study of the branes in $D=11$
theory and their implications \cite{mjd}.

An interesting and important question, which begs an 
answer, is whether the fundamental 
(super)membrane can provide the degrees of freedoms of M-theory. It is not
clear at this moment whether the quantum mechanical (super)membrane theory is a
consistent one \cite{nicol}. Thus, if the quantum theory is inconsistent,  
the answer to
the above question is negative. We may mention that not too much is 
definitively known about the consistency or inconsistency of the membrane 
theory since, as is well known, this issue is very intimately connected with
large the $N$ behaviour of the $U(N)$ matrix model \cite{gj}. 
In the recent past the proposal
of the M(atrix) model \cite{matrix} has led to very interesting developments 
\cite{rev2}.
The M(atrix) theory reveals various salient features of the M-theory. According
to the basic postulate of the M(atrix) theory, the dynamics of the eleven
dimensional M-theory finds its description in the many body quantum mechanics
of $N$ $D0$-branes of the type IIA theory in the limit $N \rightarrow \infty$.
The compactified M(atrix) theory has close connections with super
Yang-Mills theories through dualities. Furthermore, it 
provides a theoretical basis to the 
understanding of the microscopic dynamics of M-theory and holds the  promise of
exploring various aspects of string theories nonperturbatively \cite{revm}.

Recently, there have been attempts 
to study the supermembrane action in curved backgrounds with target space tensor
fields \cite{dpp}   
since, in spite of the above mentioned shortcomings, the membrane theory
does provide an intricate relation with M-theory. In this context, there
have been attempts to unravel how much the world volume theories know about
the spacetime \cite{pk2}.

The purpose of the present  investigation is to study properties of the bosonic
membrane worldvolume theory and try to expose how the theory  is encoded
with target space local symmetries, which are associated with general coordinate
transformation (CGT) 
and the gauge symmetries of the three index tensor field. We adopt 
a procedure similar to the one proposed by Veneziano \cite{g}, in the context of 
string theory, to derive  gravitational Ward identities, and by Veneziano and 
the author \cite{mv,mv1} 
  to derive  Ward identities for various massless excitations of
strings: both compactified and noncompactified \cite{m}. To briefly recall the
formalism, a Hamiltonian phase space framework is adopted and the 
Hamiltonian form of action is derived. Then, a set of generating functionals 
associated with the local (target space) symmetries of the theory are
introduced. Next, it  is shown that the variation of the action, under these
canonical transformations, can be compensated by suitable (gauge) transformation
of the massless backgrounds. Finally, it is argued that the phase space path
integral measure remains invariant, at least classically; the desired
Ward identities follow as immediate consequences. 
One of the fascinating, aspects of
our works was that the canonical transformations implemented on the worldsheet
action, indeed revealed the local symmetries of the theory in the target space.
We have adopted a similar approach, and we will show that it is possible
to introduce canonical transformations associated with general coordinate
transformation and gauge transformation in the target space of the M-theory.  
Our results are to be understood as classical one in view of the preceding
remarks regarding the quantum theory of membranes.

The bosonic  
membrane action in curved space in the presence of antisymmetric tensor
field has the following form:   
\bea S_M = T 
\int d^{3}\xi \bigg\{ \sqrt { g} 
- {1\over 6}\epsilon^{ijk}A_{ijk} 
\bigg\}, \eea
where $g_{ij}$ and $A_{ijk}$  are pullbacks to the world volume of the 
spacetime metric and the antisymmetric tensor field of eleven dimensional
supergravity; although we focus our attention on the bosonic theory
we often refer to it as $D=11$ supergravity and 
$T$ stands for the constant tension. 
When  writtten  explicitly, with ${ g} ={ \rm det }~g_{ij}$:
\bea g_{ij}= \partial _iX^M \partial _j X^N G_{MN} \eea
\bea A_{ijk} = A_{MNP}\partial _iX^M \partial _jX^N \partial _kX^P.  \eea
Here, and everywhere, lower-case latin letters $i,j,k,..$ etc., and upper-case 
letters  $M,N,P,..$etc., refer to the worldvolume and target space
indices, respectively. 
 
 It is useful to deal with a different form of action, introduced by Bergshoeff,
London and Townsend \cite{blt}. 
In this reformulation of the action, the tension of the
membrane could  be generated through the introduction of a worldvolume 
two-form potential. 
\bea S = \int d^3\xi{1\over {\lambda}} \bigg\{{\rm det}~ g +(* {\cal G})^2 
\bigg\}, 
\eea
where, $\lambda$  is a Lagrangian multiplier field.
\bea {\cal G}_{ijk}= \partial _iU_{jk}+\partial _kU_{ij}+\partial _jU_{ki} -
A_{ijk} \eea
$U_{ij} $ is the worldvolume antisymmetric tensor field, which has no local 
degrees of freedom, $g_{ij}$ and  $A_{ijk}$ are  defined above, and the dual, 
$*{\cal G} = {1\over 3} \epsilon^{ijk}
{\cal G}_{ijk}$.

The equation of motion for U, in form notation, is
${\rm d}({{* {\cal G}\over {\lambda}}}) = 0 $, and one solves   

\bea {*\cal G} = T\lambda , \eea
$T$ being a constant, to be identified as the tension. Now, if one writes down
the rest of the field equations and makes the above substitution (6), then the
bosonic membrane equations of motion are recovered. If we want to substitute
(6) into the action (4) and  derive the field equations of bosonic 
membrane, then it is necessary to add a surface term to the action. Once
the extra term is added, one gets the same result as substituting $*\cal G$
into the field equations derived from (4).

The canonical momenta $P_S$ associated with the coordinates $X^S$
are  given by 
\bea P_S = {1\over {2\lambda}} {\partial g \over {\partial {\dot X}^S}} -
{{* \cal G}\over {\lambda}} A_{SMN} \partial _iX^M \partial _jX^N 
\epsilon ^{0ij}.
\eea
We denote the worldvolume coordinates as $\tau,\xi$ and $\sigma$,
 and ${\dot X}^S
$ denotes derivative with respect to $\tau$. The derivatives with respect
to the other two worldvolume coordinates will be denoted, sometimes, as
$X^S,_i$ and $i= \xi , \sigma$ as an economy in notation. A lengthy, but
straightforward computation reveals that the 
following two relations are satisfied:
\bea P_S X^S,_i =0 \eea
and it is easy to see that these two are analogous  to the constraint 
$P\cdot X' =0$
in string theory (in that case prime was derivative with respect to $\sigma$).
The canonical momentum for the world volume gauge field is
\bea {\cal P}_{mn}= { *{\cal G}\epsilon _{0mn} \over {\lambda}}. \eea
Furthermore,
\bea {\cal H} = (P_S +{\cal P}^{mn}{\cal A}_{Smn})^2 +
 \{(X,_{\xi}^M)^2(X,_{\sigma}^N)^2
-(X,_{\xi}.  X,_{\sigma})^2 \}({\cal P}^{mn}\epsilon_{0mn})^2 =0. \eea
We identify this with the Hamiltonian density and 
 ${\cal A}_{Smn} = A_{SMN}\partial _mX^M \partial _nX^N $.
The tensor ${\cal P}_{mn}$ is antisymmetric and note that conjugate momentum of 
$U_{0n}$, ${\cal P}_{0n} =0$, just as in the case of Maxwell 
electrodynamics, the conjugate momentum of $A_0$ vanishes. Then, if one adopts
the constrained Hamiltonian approach, the Gauss law $\partial _i E^i =0$
appears as a secondary constraint. For the case at hand, the corresponding 
procedure yields  the constraint $\partial _m {\cal P}^{mn}
=0$ for the worldvolume  index antisymmetric gauge field.
The antisymmetry property of ${\cal P}_{mn}$, together with the constraint,
implies that it is proportional to $\epsilon _{mn}$.

Let us introduce a generator
of infinitesimal canonical transformation, in order to expose that the theory
is encoded with information on general coordinate invariance in target space
\bea \Phi _G = \int d\xi d\sigma P_L{\Lambda}^L(X), \eea
 $\Lambda ^L(X)$ being the infinitesimal parameter.
The coordinate and conjugate  momenta transform as follows under $\Phi _G$ 
(recall that variation of a function of phase space
variables, $F(X,P)$, is $\delta F=\left [F, \Phi _G \right]_{PB} $): 
\bea \delta _{\Phi} X^S=\Lambda ^S(X),
~~~~ \delta _{\Phi} P_L=-P_N\Lambda ^N,_{L}(X). 
\eea
Indeed, under $\Phi _G$ the coordinate $X^S$ is shifted infinitesimally, as is
the case under general coordinate transformation with $\Lambda ^S(X)$ as the
parameter.
The metric and the tensor fields are functions of the coordinates $\{X^N\}$ and
their  variations are 
\bea \delta _{\Phi} G_{MN}={\delta G_{MN} \over {\delta X^L}}\delta _{\Phi}X^L
,~~~ \delta _{\Phi} A_{MNP}={\delta A_{MNP}\over {\delta X^L}}\delta _{\Phi}X^L
,\eea
due to the shift in $X^S$.
The variation of the Hamiltonian action $S_H = \int d^3\xi \{P_S.{\dot X}^S
-H\}$ can be carried out with the above transformation rules for the 
coordinates, canonical momenta and the backgrounds. Next, we consider the
transformation of the backgrounds under general coordinate transformations
(GCT) with the following rules (treating them as tensors in the target space):
\bea \delta _{GCT} G_{ST}=-G_{SR}\Lambda ^R,_T - G_{RT}\Lambda ^R,_S - G_{ST},_R
\Lambda ^R \eea
and
\bea \delta _{GCT} A_{TUV}= -A_{RUV}\Lambda ^R,_T-A_{TRV}\Lambda ^R,_U-A_{TUR}
\Lambda
^R,_V - A_{TUV},_R\Lambda ^R. \eea

Finally, it can be shown that the following relation 
\bea \delta _{\Phi} S_H = - \delta _{GCT} S_H \eea
hold for the Hamiltonian action. Thus if we, formally,  define 
\bea {\bf Z}\left[G,A\right] = \int \left[{\rm phase~ space, ...}\right]~
{\rm exp}(-i S_H). \eea
Then we can argue that under the canonical transformation the phase space 
measure remains invariant (at least classically) and the variation of the 
Hamiltonian action under the canonical transformation $\Phi _G$ can be 
compensated by the general coordinate transformation of the backgrounds. 
Therefore, $\bf Z$ satisfies the following relation 
\bea {\bf Z}\left[G,A\right]= {\bf Z}\left [G+\delta _{GCT} G , A+\delta _{GCT} 
A\right] \eea
leading to the equation
\bea \int dx^M \bigg \{{\delta {\bf Z}\over {\delta G_{NP}(x)}}
\delta _{GCT}G_{NP}(x)
+ {\delta {\bf Z}\over {\delta A_{NPQ}(x)}} \delta _{GCT}A_{NPQ}(x) \bigg \}=0 
.\eea 
Since the infinitesimal parameter $\Lambda ^L(Y)$ is arbitrary, we can 
functionally differentiate the above equation with repect $\Lambda ^L(Y)$ to 
arrive at 
\bea \int d^Mx\bigg \{ {\delta {\bf Z}\over {\delta G_{MN}(x)}}\left[G_{ML}(x)
\partial _N\delta(x-Y)+G_{LN}(x)\partial _M\delta (x-Y)\right. \nonumber \\
+ \left. \partial _L G_{MN}(x)
\delta (x-Y) \right]
+  {\delta {\bf Z}\over {\delta A_{MNP}(x)}}\left[  A_{LNP}\partial _M
\delta (x-Y)+A_{MLP}\partial _N\delta (x-Y)\right. \nonumber \\
+ \left. A_{MNL}(x)\partial _P\delta (x-Y)
+\partial _LAMNP \delta (x-Y)\right]\bigg \} =0.  \eea
Thus we see that ${\bf Z}$ exhibits invariance under general coordinate 
transformations, a result similar to the property of the generating functional
considered by us  in the case of string theory by  us \cite{mv,mv1}. 

A few remarks are in order at this stage. The action, $S_H$, 
appearing in eq.(17)
is the Hamiltonian action and it contains the other pieces such as the 
ghost part (in case of covariant quantization), 
 a piece taking into account the constraints, etc. Explicit checks show
that taking constraints (8) into account does not change the relation (16).  
 We are fully aware that 
constrained Hamiltonian \cite{hrt} BRST quantization of 
higher dimensional extended objects, along the line
one adopts for string theory, is difficult to accomplish because of  various
technical hudles  that one encounters \cite{h}.  
However, if the additional pieces in the action 
do not depend on the phase space variables $\{X^L, P_L\}$, then under
the canonical transformation (12) which is induced by the generator $\Phi _G$, 
 the relation (16) will be  
satisfied since it is a functional of only $\{X,P\}$. However, it is not
necessary to work in the frame work of covariant formulation to derive the
Ward identities. One could make a suitable gauge choice, as discusses in 
\cite{dpp} following the work of \cite{mgjh} and one will be able to discuss
the local symmetries following the approach of Veneziano \cite{g}
as he obtained the
Ward identities without resorting to BRST formalism. It is worth mentioning
that in the Hamiltonian phase space approach the results are derived
elegantly.
  We might repeat that our purpose here is to explore how much information  
about  the local target space symmetries can be extracted from point of view 
of the worldvolume action (4) of the membrane. 

In order to explore the gauge symmetry associated with the three
index antisymmetric tensor field, let us introduce another generator
for the canonical transformation: 
\bea \Gamma _A = \int d\xi d\sigma {\cal P}^{mn} {\Psi}_{AB}(X)\partial _m X^A
\partial _nX^B \eea
Here $m, n= \sigma , \xi$ components tensor. 
We find  that $\delta _{\Gamma} X^A =0$, and
\bea \delta _{\Gamma} P_A= -{\cal P}^{mn}\left[\partial _A\Psi _{BC}\partial _m
X^B\partial _nX^C+\partial _C\Psi _{AB}\partial _mX^C\partial _nX^B+\partial _B
\Psi _{CA}\partial _mX^C\partial _nX^B \right]. \eea
We can  thus compute the variation of $S_H$ from the above
transformation rules. Note that $\delta _{\Gamma} G_{MN} =0$ and $\delta _{
\Gamma} A_{MNP}=0$ since  $X^A$'s  have vanishing shift 
under $\Gamma _A$. 

The variation of the antisymmetric tensor field,  under the local gauge
transformation, is given by
\bea \delta _{gauge} A(X)_{BCD} = \partial _B\Psi (X) _{CD}+\partial _D
\Psi (X)_{BC}+
\partial (X)_C\Psi _{DB}. \eea

After some long, but straightforward
calculations, we find that  
\bea \delta _{\Gamma} S_H = -\delta _{gauge} S_H \eea
is satisfied. If we adopt  arguments similar to the 
one for deriving the invariance
properties of the generating functional under GCT, 
the corresponding relation is
\bea \int d^Mx {\delta {\bf Z}\over {\delta A(x)_{BCD}}}
\delta _{gauge} A(x)_{BCD}=0.
\eea
Recall that $\delta _{gauge} A_{BCD}$, given by (23), involves the 
gauge parameters $\Psi _{AB}$. Therefore, if we  
 functionally differentiate eq. (25) with respect to $\Psi _{AB}(Y)$, the  
final expression is
\bea \int d^Mx {\delta {\bf Z}\over {\delta A_{BCD}}}\left[\partial _B\delta(x
-Y)\delta ^N_C\delta ^P_D+\partial _D\delta (x-Y)\delta ^N_B\delta ^P_C+
\partial _C\delta (x-Y)\delta ^N_D\delta ^P_B \right]=0 .\eea
This is the gauge invariance property of the generating functional $\bf Z$ 
and a similar relation was derived in the 
context of string theory \cite{mv,mv1} 
 to obtain the gauge Ward
identities associated with the two form potential. 
A natural next step would have been to take  a string
 functional derivative of
eq. (20) and eq. (26) with respect to the background fields $G(Y_i)_{MN}$ and 
$A(Y_i)_{MNP}$.
Notice that the right-hand sides of both these equations will still be 
zero. In the case of string theory, it was possible to test
such Ward identities (modulo anomalies) 
for choice of simple background configurations. However, on this occasion, 
although we demonstrate the invariance properties of the generating functional,
we are not in a position to carry out  explicit  computations. It is worth while
to mention that NS-NS branes and D-branes, 
in the context of type IIA and IIB theories, 
appear as extended objects and it might be possible
to check our results, for the corresponding antisymmetric fields, 
through explicit calculations. 
This might be achieved by adopting a more powerful technique as was utilized to
explore gauge symmetries \cite{mm} associated with the excited levels 
of string states explored earlier \cite{more}. We
hope to report our results in these directions in the future.

We present below some results on type the IIB string action and
study a few interesting  properties of the action in the light 
of our investigations
of the worldvolume action. This is intimately connected with the works of
Townsend and of Cederwall and Townsend \cite{ct}.

In order to construct a manifestly S-dual type IIB superstring action, let us 
recall that type IIB string is endowed with a pair of two form tensor fields
coming from the NS-NS and R-R sectors. Therefore, a pair of worldsheet gauge
potentials, denoted by $V_i$ and ${\tilde V}_j$, are introduced, and the 
corresonding modified field strengths are
\bea F_{ij}=\partial _iV_j-\partial _jV_i -B_{ij},~~ {\tilde F}_{ij}=\partial
_i{\tilde V}_j -\partial _j{\tilde V}_j -{\tilde B}_{ij},  \eea
where  $B_{ij}=B_{MN}\partial _iX^M\partial _jX^N$ and ${\tilde B}_{ij}=
{\tilde B}_{MN}\partial _iX^M\partial _jX^N$ are the pullbacks of the two
antisymmetric tensor fields coming from the NS-NS and the 
R-R sectors respectively.
We note that scalar dilaton, $\phi$, and pseudoscalar axion, $\chi$, coming from
the NS-NS and the R-R sectors, are also
present in the massless spectrum of type IIB theory. The manifest $SL(2,Z)$ 
 invariant action is
\bea S= {1\over 2}\int d^2\xi{\lambda}\bigg \{ g + e^{-\phi}(*F)^2 +e^{\phi}
\left[*({\tilde F} -\chi F)^2\right] \bigg \}, \eea
where $\lambda$ is  again  the Lagrange multiplier field, $*F=\epsilon ^{ij}
F_{ij}$ and $*{\tilde F}=\epsilon ^{ij}{\tilde F}_{ij}$ are the 
worldsheet scalar densities, $g_{ij}=g_{MN}\partial _iX^M\partial _jX^N$ is the
pullback of the `Einstein frame' metric and $ g= {\rm det }~ g_{ij}$. Notice 
that the dilaton and axion can be combined to form  the $SL(2,R)$ matrix
and $B_{MN}$ and ${\tilde B}_{MN}$ identified as a doublet  
 as follows:
\bea 
{\cal S}\equiv \left( \begin{array}{cc} \chi ^2e^{\phi}+e^{-\phi} & \chi 
e^{\phi} \\ \chi e^{\phi} & e^{\phi} \end{array} \right),~~~~
{\hat {\bf B}}_{MN} \equiv \left( \begin{array}{cc} B_{MN} \\ {\tilde B}_{MN}
\end{array} \right). \eea
Under $SL(2,Z)$, ${\cal S} \rightarrow {\cal U}{\cal S}{\cal U}^T $ and
${\hat {\bf B}} \rightarrow ({\cal U}^T)^{-1} {\hat {\bf B}}, ~{\cal U}\in 
SL(2,Z)$. The 
worldsheet gauge fields $V_i$ and ${\tilde V}_j$ can be paired in
the same way as
the 2-form potentials and their transformation laws  are required to be exactly
the same as those  of the two-form potentials to ensure $SL(2,Z)$ 
invariance of the action.
The equations of motion of ${\tilde V}$ lead to the relation:
\bea *{\tilde F}=e^{-\phi} \lambda T_s ,\eea
$T_s$ is a constant and is identified to be the tension. 
It is easy to derive the Hamiltonian constraint
\bea H= (P_M+{\tilde E}{\tilde B}_M+EB_M)^2+(X')^2\left[e^{\phi}(E+\chi 
{\tilde E})^2 +e^{-\phi}{\tilde E}^2 \right] \eea
and the one generating the  $\sigma$ reparametrization corresponds to $P_M{X'}
^M$; here $\tau , \sigma $ denote the worldsheet coordinates; $P_M$ is
conjugate momentum of $X^M$; $E$ and ${\tilde E}$, are 
momenta conjugate to
$\sigma$ components of potentials $V$ and ${\tilde V}$ respectively 
and the prime denotes 
derivative with respect to $\sigma$. For brevity of notation,
\bea B_M={X'}^NB_{MN},~~~~~ {\tilde B}_M={X'}^N{\tilde B}_{MN}. \eea
As is well known, in constraint analysis, we shall end up with the Gauss law
conditions $\partial _{\sigma} E=0$ and $\partial _{\sigma}{\tilde E}=0$,
 leading
to the conclusion that $E$ and ${\tilde E}$ are independent of $\sigma$. 
Furthermore, the field equations for $V_{\sigma}$ and ${\tilde V}_{\sigma}$
components imply that $E$ and ${\tilde E}$ do not vary with time and these 
electric fields take integer values in appropriate units. Thus the relevant
Hamiltonian action is
\bea S_H=\int d^2\xi \left[{\dot X}^MP_M -{1\over 2}{\lambda}\bigg\{(P_M+m
{\tilde B}_M +nB_M)^2 +{X'}^2\left[ e^{\phi}(m+n\chi)^2+e^{-\phi}n^2\right]
\bigg\} \right]. \eea
Here $dot$ stands for $\tau$ derivative and it is understood that we also add
the constraint $X'\cdot P$ with its multiplier to the above equation.

We want to demonstrate how the gauge invariances of this theory, associated
with the pair of two-form potentials, appears in the approach we have been
persuing. The generator associated with the NS-NS, B-field 
canonical transformation
is 
\bea \Phi _B= E\int d\sigma \Psi _M(X){X'}^M ;\eea
with the other one, it is
\bea \Phi _{\tilde B}= {\tilde E}\int d\sigma {\tilde \Psi}_M(X){X'}^M .\eea
Note the appearance of constant electric fields (which are quantized) 
explicitly in the above formulae. As before, one can compute the variations of
all the phase space variables under $\Phi _B$ or $\Phi _{\tilde B}$ and, if so
desired, one can compute the variations when both the canonical 
transformations are implemented. Also note that the variation of the two 
backgrounds, under the Abelian gauge transformations, are
\bea \delta _{gauge}B = \partial _M\Psi _N - \partial _N\Psi _M,~~\delta _
{gauge'}{\tilde B}=\partial _M{\tilde \Psi} _N
- \partial _N{\tilde \Psi} _M .\eea
Then, one arrives at the following result
\bea
\delta _{\Phi _B + \Phi _{\tilde B}} S_H = (\delta _{gauge} +\delta _{gauge'})
S_H ,\eea
which eventually leads to the Ward identities derived by us \cite{mv,mv1}
in the context
of bosonic strings (both compact and noncompact). Neeedless to say that, with
a generating functional $\int d\sigma P_M\Lambda ^M(X)$, we can derive the
gravitational Ward identities in a straightforward manner.

It has been argued that the type IIB action also provides  glimpses of the
12-dimensional world, once the coordinates are  identified  as 
${\bf X}=(X^M, V_1, {\tilde V}_1)$,
 the corresponding canonical momenta as ${\bf P}=(P_M, E,{\tilde E})$,
and then the Hamiltonian action could be written in a compact form. Although
the constraints are not invariant under 12 dimensional Lorentz transformation,
there is invariance under ten dimensional Lorentz transformation and the 
$SL(2,R)$. We would like to present an interesting form for the Hamiltonian,
which is strikingly similar to the one derived by us for strings in the presence
of backgrounds and there is a matrix which is 
very much like the {\bf M}-matrix
used in the context of $O(d,d)$ symmetry. The Hamiltonian can be written as
\bea {\cal H} = {1\over 2}{\cal P}^T {\cal M}{\cal P} \eea
in a compact matrix notation, with the following definitions:
\bea {\cal P} = \left( \begin{array}{cc}{\bf P}_M & {{\bf X}'}^M \end{array} 
\right) \eea
\bea {\cal M} = \left( \begin{array}{cc} g^{MN} & g^{MP}(EB_{PN}+{\tilde E}
{\tilde B}_{PN})
 \\
-(EB_{MP}+{\tilde E}{\tilde B}_{MP})g^{PN} &
{\cal W}^2g_{MN}+\left[2E{\tilde E}Bg^{-1}{\tilde B}+E^2
Bg^{-1}B+ {\tilde E}^2 {\tilde B}g^{-1}{\tilde B} \right]_{MN} \end{array}
\right)  \eea 
and
\bea {\cal W}^2= e^{\phi}(E+\chi {\tilde E})^2+e^{-\phi}{\tilde E}^2
.\eea
 The structure of ${\cal W}$ is quite interesting:  from the S-duality
attributes of the type IIB theory,
the combination of dilaton and axion tells us about the tension. If we go to
the string frame with the definition $g_{string}=e^{\phi}$, tension is given by 
$ T_s=\sqrt {{{n^2}\over {g^2_{string}}} +(m+n\chi)^2}$,  
in agreement with the results of Schwarz \cite{john}. On the other hand, 
when we envisage IIB
theory from the F-theory perspective \cite{vb}, 
$\phi$ and $\chi$  are the modular parameters
of the torus and are geometrical objects from the  point of view of a 
12-dimensional theory. Furthermore, the pair of integers $(m,n)$ 
have a simple interpretation as quantized momenta.

In summary, we considered the membrane action [19] 
in curved space in the presence 
of the antisymmetric tensor field. The action has a novel property that 
a worldvolume gauge field and a Lagrange multiplier field are  introduced;
when we  solve  the equations of motion of the gauge field on the world volume
the tension appears (see eq. (6)). 
We introduced generators of canonical transformations
associated with general coordinate transformations and the gauge transformation
of the three index antisymmetric tensor field. Then using the techniques 
introduced by Veneziano and by Veneziano and me, the invariance properties of
the generating function are derived. These results are to be envisaged
as classical ones, since  anomalies might afflict these results owing  to  
quantum effects in some cases. Next, we considered  a manifestly 
S-duality invariant 
action for type IIB theory and showed how the target space local symmetries
can be exhibited by  introducing generators of canonical transformation
leading to invariance of the generating function. We point out how the
Hamiltonian of the type IIB action, in the 
presence of the worldsheet gauge field
resembles  the well known M-matrix (which very often appears in the context of
$O(d,d)$ symmetry) when we perceive the Hamiltonian from the F theory
perspective. Our approach might be useful to provide further understanding
of the properties of the worldvolume theory of D-branes.

\noindent{\bf Acknowledgements}:
It is a pleasure to thank Gabriele Veneziano for very valuable discussions and
for his encouragements. I have benefited from  useful discussions 
with  Hermann Nicolai.

\end{document}